\documentclass[preprint2]{aastex}

\usepackage{graphicx}
\usepackage{subcaption}
\usepackage{color}

\shorttitle{A Multi-Period Oscillation in a Stellar Superflare}
\shortauthors{Pugh, Nakariakov \& Broomhall}

\begin{document}

\title{A Multi-Period Oscillation in a Stellar Superflare}

\author{C. E. Pugh\altaffilmark{1}, V. M. Nakariakov\altaffilmark{1, 2} and A.-M. Broomhall\altaffilmark{1, 3}}

\altaffiltext{1}{Centre for Fusion, Space, and Astrophysics, Department of Physics, University of Warwick, Coventry, CV4 7AL, UK; c.e.pugh@warwick.ac.uk}
\altaffiltext{2}{Central Astronomical Observatory at Pulkovo of RAS, St Petersburg 196140, Russia}
\altaffiltext{3}{Institute of Advanced Study, University of Warwick, Coventry, CV4 7HS, UK}

\begin{abstract}
Flares that are orders of magnitude larger than the most energetic solar flares are routinely observed on Sun-like stars, raising the question of whether the same physical processes are responsible for both solar and stellar flares. In this letter we present a white-light stellar superflare on the star KIC9655129, observed by NASA's \emph{Kepler} mission, with a rare multi-period quasi-periodic pulsation (QPP) pattern. Two significant periodic processes were detected using the wavelet and autocorrelation techniques, with periods of $78 \pm 12$\,min and $32 \pm 2$\,min. By comparing the phases and decay times of the two periodicities, the QPP signal was found to most likely be linear, suggesting that the two periodicities are independent, possibly corresponding either to different magnetohydrodynamic modes of the flaring region, or different spatial harmonics of the same mode. The presence of multiple periodicities is a good indication that the QPPs were caused by magnetohydrodynamic oscillations, and suggests that the physical processes in operation during stellar flares could be the same as those in solar flares.
\end{abstract}


\section{INTRODUCTION}

Solar flares typically release $10^{29}$ to $10^{32}$\,erg of energy over a timescale of up to several hours. Many stars not dissimilar from the Sun have been observed to produce flares several orders of magnitude more powerful than any solar flare on record, with amplitudes around 0.1--1\% of the stellar luminosity and estimated energies of $10^{33}$ to $10^{36}$\,erg \citep{2012Natur.485..478M}. Due to the potential for substantial damage in the near-Earth environment associated with powerful flares, it is important to work towards determining whether a superflare could occur on the Sun, and if so what the probability of one occurring at a given time would be. \citet{2013PASJ...65...49S} found that, in principle, a $10^{34}$\,erg superflare could occur on the Sun if the necessary magnetic flux were stored over one solar cycle period ($\sim$11\,yrs), while it would take around 40\,yrs to accumulate enough magnetic flux for a $10^{35}$\,erg superflare. It is not completely understood how magnetic flux could be prevented from emerging from the base of the convection zone for such a long period of time, hence additional observational data and theoretical constraints are needed to support or disprove these findings.

Quasi-periodic pulsations (QPPs) are time variations in the intensity of light emitted by a flare. They have been widely observed in solar flares \citep[e.g.][]{2010SoPh..267..329K}, and since they are a common feature of flares, QPPs can be used to constrain parameters of the plasma in the flaring region and indicate the physical processes in operation. The specific mechanism responsible for QPPs is unknown, but several possible mechanisms have been proposed. These fall into one of two categories: `load/unload' (or `magnetic dripping') mechanisms, or magnetohydrodynamic (MHD) oscillations \citep[e.g.][]{2009SSRv..149..119N, 2010PPCF...52l4009N}. The load/unload mechanisms could apply where there is a continuous supply of magnetic energy, causing magnetic reconnection to repetitively occur each time a threshold energy is surpassed. In this scenario, the periodicity is connected with a quasi-periodic self-oscillatory regime of spontaneous magnetic reconnection, which would result in quasi-periodic modulation of particle acceleration and the rate of energy release. Examples of this regime have been found in numerical simulations by \citet{2009A&A...494..329M, 2000A&A...360..715K, 2012A&A...548A..98M}. QPPs caused by MHD oscillations could involve either oscillations of the flaring region itself, or MHD oscillations of a nearby structure. In the first case, variation of parameters of the flaring plasma (such as the magnetic field and plasma density) could directly modulate the radiation emission, or could result in periodic modulation of particle acceleration and hence emission via the gyrosynchrotron mechanism or bremsstrahlung. The latter case may be considered as a periodically triggered regime of magnetic reconnection, where the fast or slow magnetoacoustic oscillations leak from the oscillating structure and approach the flaring site, resulting in, for example, plasma micro-instabilities and hence anomalous resistivity, which could trigger reconnection in the flaring loop \citep{2006A&A...452..343N, 2006SoPh..238..313C}.

The first observation of oscillations in a stellar flare was made by \citet{1974A&A....32..337R}. Since then occasional observations of QPPs in different stars have been made in the optical \citep{2003A&A...403.1101M}, ultraviolet \citep{2006A&A...458..921W}, microwave \citep{2004AstL...30..319Z} and X-ray \citep{2005A&A...436.1041M} wavebands. Recently, \citet{2013ApJ...773..156A} found QPPs in a megaflare on the dM4.5e star YZ CMi, observed in white light, which looked very similar to oscillations in solar flares that were concluded to be the result of standing longitudinal modes. This suggests that this mechanism, where the oscillations cause plasma parameters to vary and hence modulate the acceleration of precipitating non-thermal electrons, applies for a wide range of flare energies, including superflares. So far no other evidence has been found to suggest any major differences between solar and stellar QPPs, indicating that the basic physical processes responsible for the energy releases (e.g. magnetic reconnection) are the same.

NASA's \emph{Kepler} mission is proving to be an excellent resource for the study of stellar flares, owing to the large number of stars observed over long periods of time, the high sensitivity of its photometric observations, and the availability of data with a cadence of one minute: suitable for studying QPPs with periods greater than a few minutes. \citet{2015MNRAS.450..956B} found several instances of quasi-periodic variability in stellar flares using \emph{Kepler} data, the periods of which were not found to correlate with any global stellar parameter, suggesting that they could be QPPs in the impulsive energy release itself.

In this study we focus on a rare example of a flare showing evidence of two periods of oscillation, which occurred on KIC9655129, a K-type eclipsing binary star. While at this stage it is difficult to tell from which of the binary stars the flare originates, the conclusions of this paper do not rely upon the stellar spectral type. This star has a low contamination factor of 0.035, meaning that little of the light detected comes from surrounding objects. Multiple period QPPs are of great interest because they impose additional constraints on the plasma parameters, have implications for the underlying QPP mechanisms, and have interesting seismological implications \citep[e.g.][]{2007A&A...473..959V, 2009A&A...493..259I, 2011ApJ...740...90V}. Detections of multiple period QPPs are only very occasionally made in solar flares \citep[e.g.][]{2009A&A...493..259I}, owing to the lower amplitudes of higher harmonics, and they are even rarer in stellar flares, most likely due to the noise level of the data. Only two examples in stellar flares have been reported previously, both of which were detected in different wavebands to the flare presented in this paper. The first case was found in the microwave band by \citet{2004AstL...30..319Z}, which had one quasi-periodic component with a period that varied from 0.5 to 2\,s during the flare, and the other was a series of pulses with a period of 0.5\,s. The second case was found in the X-ray band by \citet{2013ApJ...778L..28S}, which had periods of 21.0 and 11.5\,min. For these two cases the detected periodicities are different to the ones found in our study, and the flares themselves may or may not be superflares, since the flare amplitudes were not specified. Moreover, a white-light emission burst is considered to be a signature of a superflare. An overview of the data and analysis methods employed to deduce the periods of oscillation are given in Section 2, while the results and details of the analysis are given in Section 3. A summary is given in Section 4. 

\section{DATA AND ANALYSIS}

\begin{figure*}
	\centering
	\begin{subfigure}[t]{0.45\textwidth}
		\centering
		\includegraphics[width=.93\linewidth]{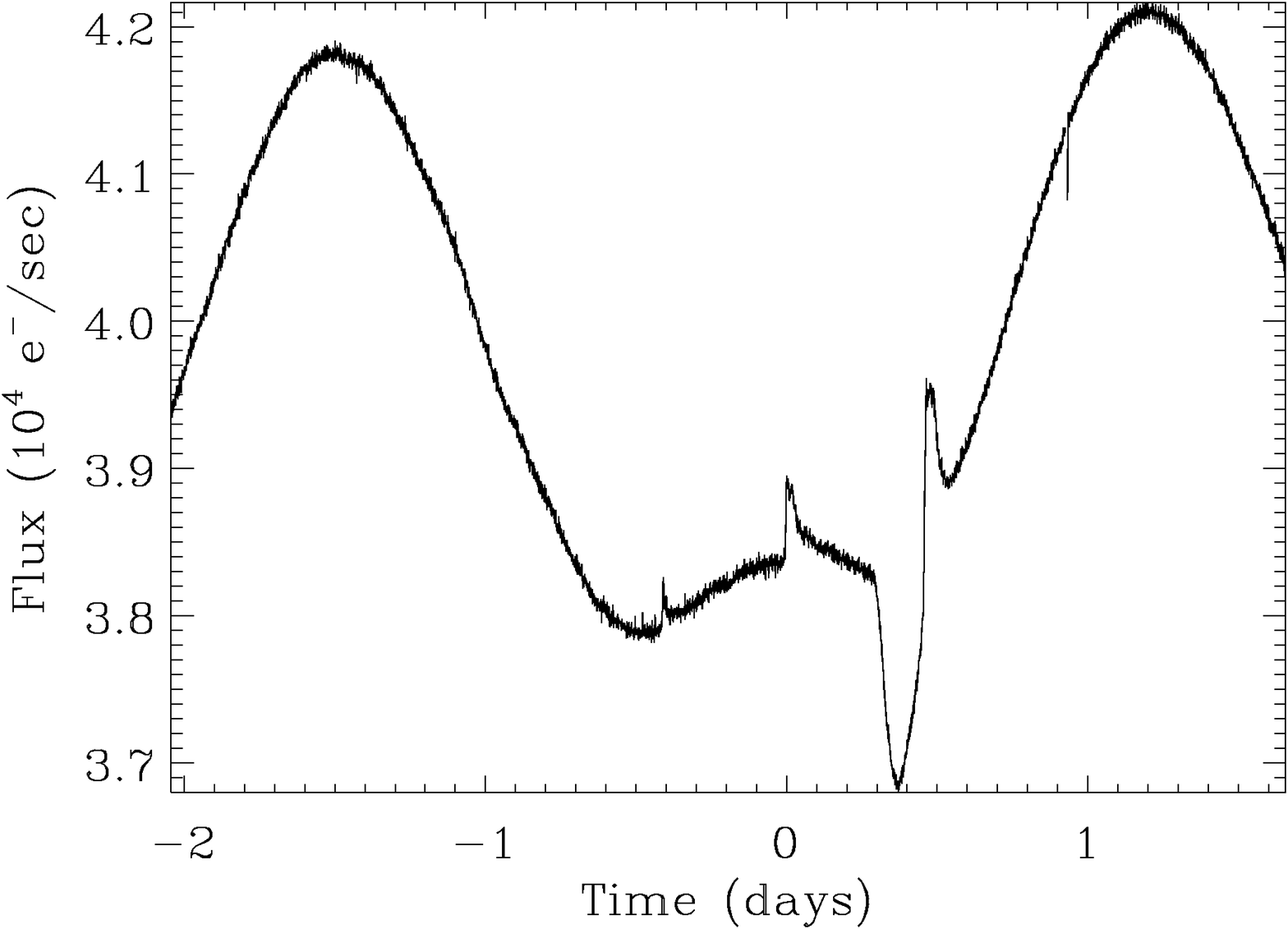}
	\end{subfigure}
	\quad
	\begin{subfigure}[t]{0.45\textwidth}
		\centering
		\includegraphics[width=.95\linewidth]{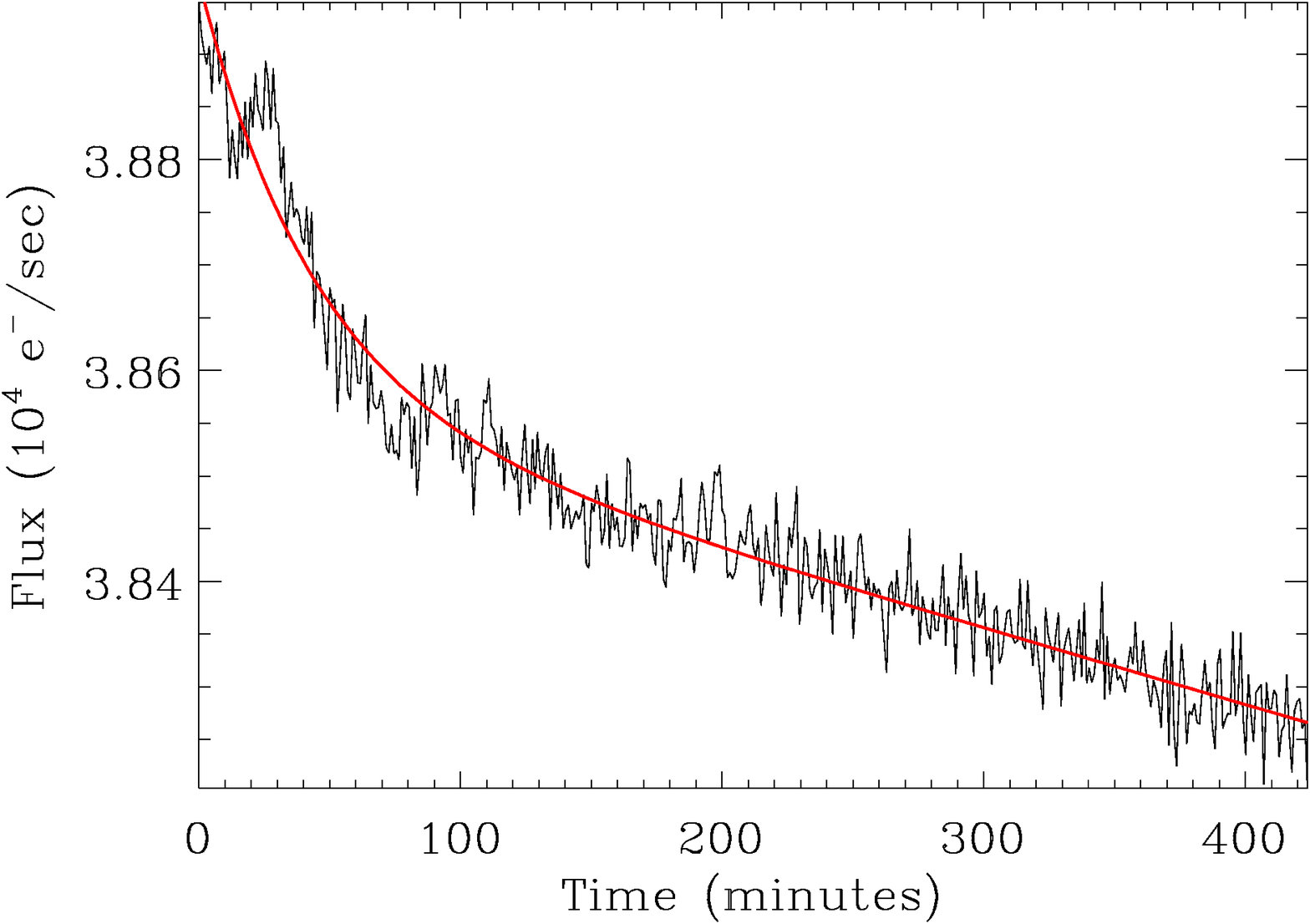}
	\end{subfigure}
	\caption{Left: A section of the short cadence light curve of KIC9655129 from Quarter 14b, which contains three flares. Right: A shorter section of the light curve, showing the decaying phase of the central flare in the plot on the left. The peak intensity of the flare is at time, $t = 0$. The red overplotted line is the result of a least-squares fit to the flare decay, as detailed in Section 3.}
	\label{fig1}
\end{figure*}

The majority of data from \emph{Kepler} have a cadence of 30 minutes, but several thousand stars have also been observed with a one minute cadence, which is more appropriate for the study of QPPs in flares. The flare of interest occurred on 2012 August 9$^{\rm th}$, and can be found in the short cadence light curve from Quarter 14b. Two types of flux data are available for each light curve: SAP and PDCSAP. The main difference is that PDCSAP data has had artifacts and systematic trends removed \citep{2012PASP..124..963K}. The pipeline module used to produce the PDCSAP flux is designed to optimise the data for the detection of exoplanet transits, however in some cases variations and artifacts that are astrophysical in nature are also removed. Fortunately, SAP flux data is suitable for the study of stellar flares, as the flares can easily be distinguished from most artifacts by comparing with the PDCSAP data, and the timescales of systematic trends are much greater than the durations of the flares.

An algorithm similar to that detailed in \citet{2011AJ....141...50W} was used to search all short cadence \emph{Kepler} data for candidate flaring A to M-type stars, and these were then checked by eye. Any flares showing potential signs of QPPs were analysed using the wavelet and autocorrelation techniques (detailed below). Of the 2,982 flares detected on 215 stars using this method, 73 showed evidence of QPPs, 11 of which had stable periodicities, and one was found with multiple significant periodicities. This is the flare studied in the remainder of this paper. A section of the light curve from KIC9655129 containing the flare of interest is shown in the left-hand panel of Figure \ref{fig1}. The periodic modulation of the light curve is due to this star being a binary, and the dip is where one star eclipses the other. The three small peaks near the centre are flares, and the right-hand panel of Figure \ref{fig1} shows the decaying phase of the central flare.
 
To help detect QPPs, the decaying phase of the flare light curve was fitted with an aperiodic analytical function using a least-squares method, and the fit was then subtracted from the light curve. Details of the fitting are given in Section 3. The resultant de-trended light curve was padded with zeros at the start, and a wavelet transform was then performed using the Morlet wavelet, to highlight any periodicities. In order to obtain estimates for the periods of the two detected periodicities, a function (described in Section 3) was fitted to the flare decay, and uncertainties were estimated using Monte Carlo simulations.

Another useful technique for reducing noise and highlighting periodicities in the data is the autocorrelation method, where the correlation of a signal with itself is calculated as a function of time lag. In this study, the time lags used ranged from zero to half the duration of the flare decay time series, with a spacing equal to the cadence of the data. The periodic variability of the autocorrelation function was then studied.

\section{RESULTS AND DISCUSSION}

In order to remove the underlying trend in the flare decay light curve, and hence emphasise any short-term variability, the following expression was fitted to the light curve, as shown by the red overplotted line in the right-hand plot of Figure \ref{fig1}:
\begin{equation}
F(t)  = Ae^{-t/t_{0}} + Bt + C,
\end{equation}
where $F$ is the flux, $t$ is time, and $A$, $t_{0}$, $B$, $C$ are constants. An exponential decay expression was chosen because, for most cases, it fits the decaying phase of the flares well. One complication is underlying trends in the light curves, which are usually due to differential velocity aberration, rotational variability (if the star is a binary or has star spots), and/or transits. The addition of a linear term, $Bt$, was enough to account for the underlying trend in the vicinity of the flare for this case.

\begin{figure}[h]
		\centering
		\includegraphics[width=0.9\linewidth]{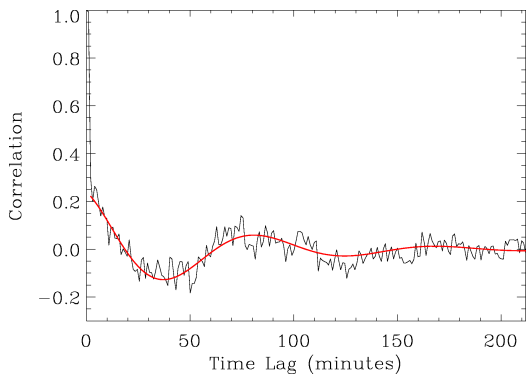}
		\caption{The autocorrelation function of the de-trended flare decay light curve (shown in Figure \ref{fig3}a), with a fitted decaying sinusoid overplotted in red, which has a period of $86.1 \pm 5.4$\,min.}
		\label{fig2}
\end{figure}

To aid in visualising the QPPs, an autocorrelation was performed on the light curve after subtracting the decay trend, as shown in Figure \ref{fig2}. The trend function in Equation 1 is aperiodic, hence its subtraction from the signal cannot introduce artificial periodicities. A decaying oscillation can clearly be seen in this plot, and the following expression, shown by the red overplotted line, fits very well:
\begin{equation}
F(t)  = Ae^{-t/t_{0}}\cos\bigg(\frac{2\pi}{P}t + \phi\bigg) + C,
\end{equation}
where $P$ is the period, $\phi$ is the phase, and $A$, $t_0$, $C$ are constants. This fit gives a period of $86.1 \pm 5.4$\,min, where the error was obtained by performing Monte Carlo simulations. 

\begin{figure*}
	\centering
	\begin{subfigure}[!ht]{0.45\textwidth}
		\centering
		\includegraphics[width=.9\linewidth]{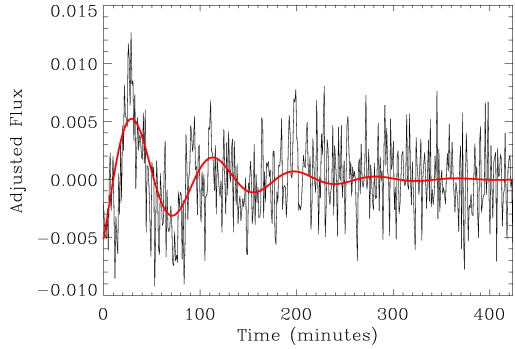}
		\caption{The de-trended decay phase of the flaring light curve, with a fit to the main periodicity overplotted in red.}
	\end{subfigure}
	\quad
	\begin{subfigure}[!ht]{0.45\textwidth}
		\vspace{-7pt}
		\centering
		\includegraphics[width=0.9\linewidth]{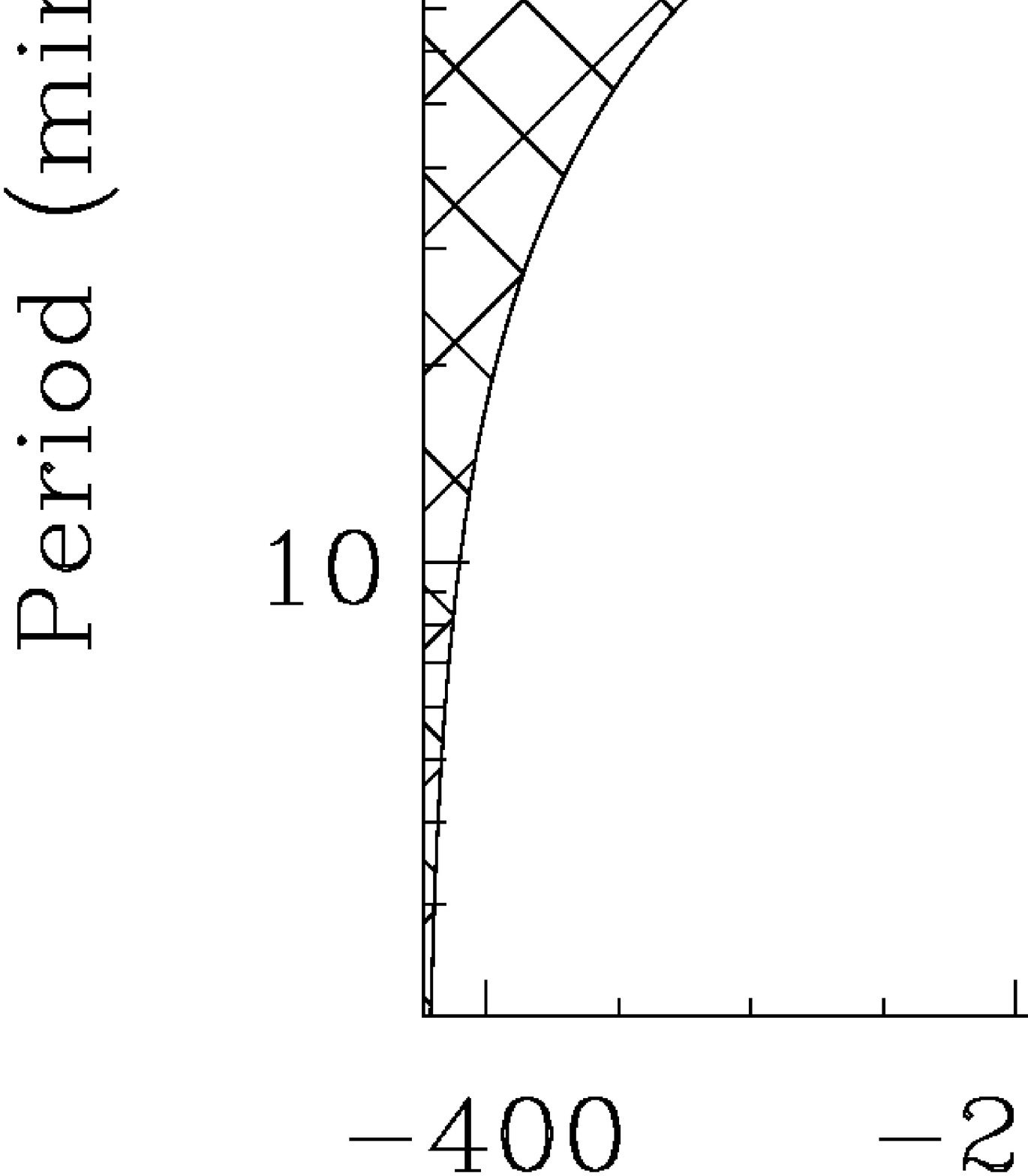}
		\vspace{+2pt}
		\caption{The wavelet spectrum of plot (a). The bright feature has a period of $84^{+25}_{-19}$\,min.}
	\end{subfigure}
	\quad
	\begin{subfigure}[!ht]{0.45\textwidth}
		\centering
		\includegraphics[width=.9\linewidth]{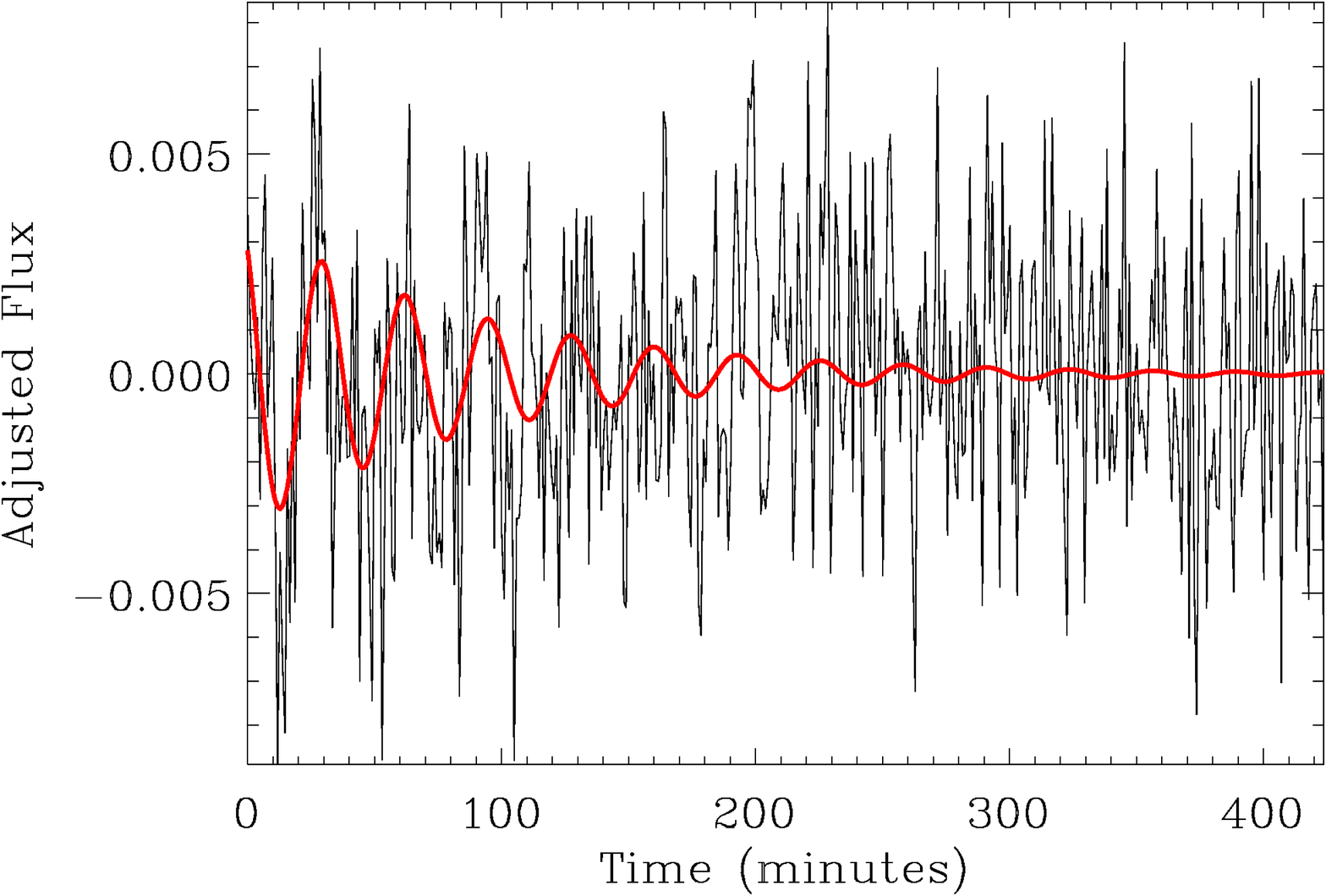}
		\caption{The de-trended light curve in plot (a) with a fit to the main periodicity subtracted. Fitting the shorter periodicity gives the curve overplotted in red.}
	\end{subfigure}
	\quad
	\begin{subfigure}[!ht]{0.45\textwidth}
	 	\vspace{-7pt}
		\centering
		\includegraphics[width=0.9\linewidth]{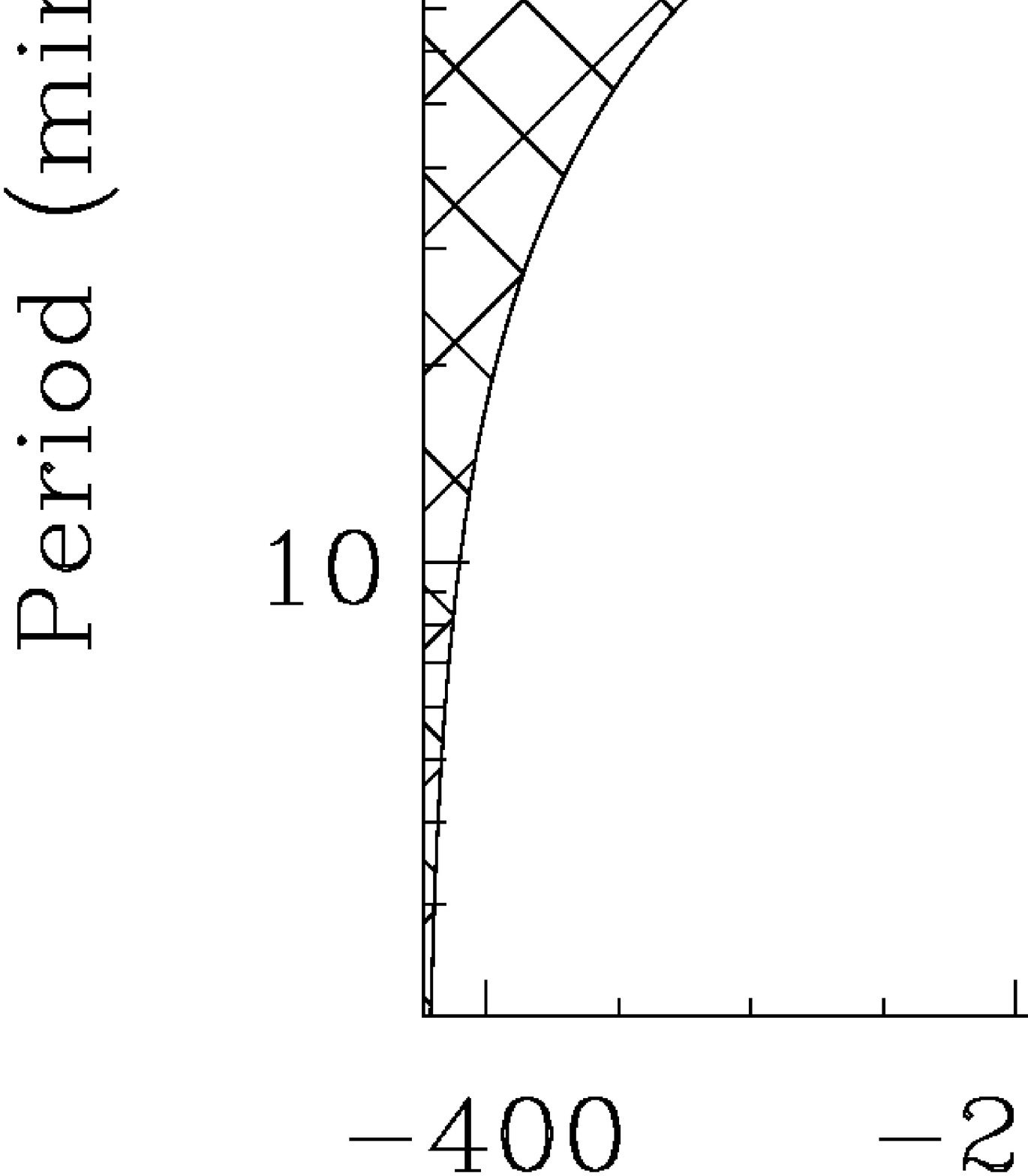}
		\vspace{+2pt}
		\caption{The wavelet spectrum of plot (e), showing a second period of $32 \pm 7$\,min.}
	\end{subfigure}
	\caption{Time series are shown on the left with their corresponding wavelet spectra on the right. The far-right panels show the global wavelet spectra. The beginnings of the time series used to produce plots (b) and (d) have been padded with zeros in order to bring the features of interest into the centre of the cone of influence. In each case the peak of the flare is at the time $t = 0$.}
	\label{fig3}
\end{figure*}

Figure \ref{fig3}a shows the de-trended light curve. The wavelet spectrum, shown in Figure \ref{fig3}b, has a clear feature at a period of $84^{+25}_{-19}$\,min, which is above the 99\% confidence level (as defined by \citealt{1998BAMS...79...61T}). There is also a feature above the 99\% level at a period of around 250\,min, which can be ignored as its duration is roughly equal to its period, so cannot be considered to be an oscillatory pattern. The small short-period features are due to noise in the data. The high-power spectral feature appears to split into two bands, suggesting the presence of two different periodicities.

To examine a possible second periodicity, we subtracted the signal given by Equation 2 with the best-fitting coefficients from the de-trended original light curve, and then performed a wavelet transform on the resultant time series. The remnant signal is shown in Figure \ref{fig3}c, with a decaying sinusoidal fit overplotted in red. Despite the noise in this plot, several cycles of the oscillation can be seen near the start, and this becomes more clear when a wavelet transform is performed, as shown in Figure \ref{fig3}d. The bright band above the 99\% confidence level shows a second period of $32 \pm 7$\,min. There is also evidence of a possible third periodicity at around 19\,min, but it requires a more complex fit.

\begin{figure}[ht]
		\centering
		\includegraphics[width=\linewidth]{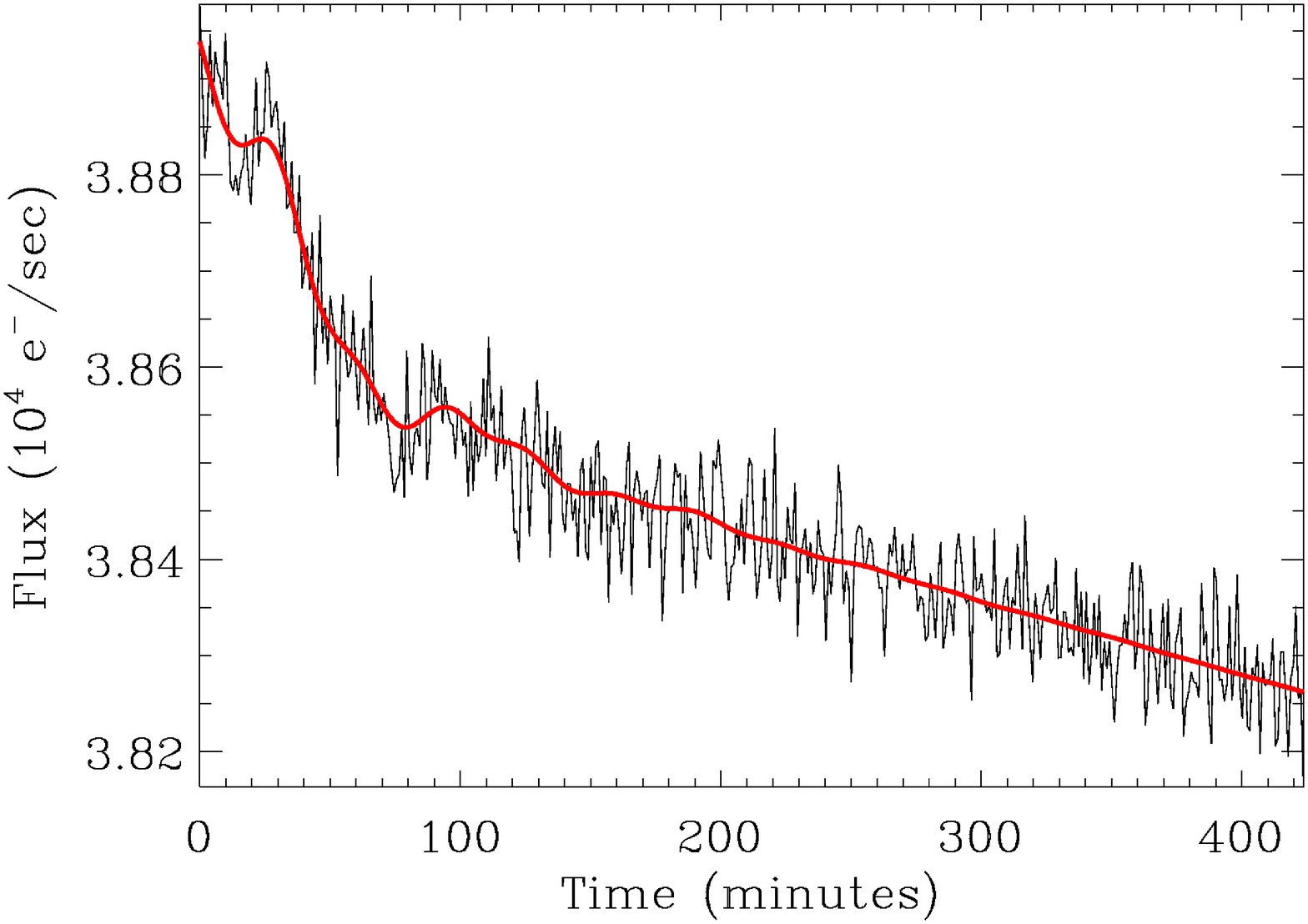}
		\caption{The flare decay light curve with the red overplotted curve showing the result of a least-squares fit to the flare decay, along with the two periodicities.}
		\label{fig4}
\end{figure}

\begin{figure*}
	\centering
	\begin{subfigure}[t]{0.4\textwidth}
		\centering
		\includegraphics[width=\linewidth]{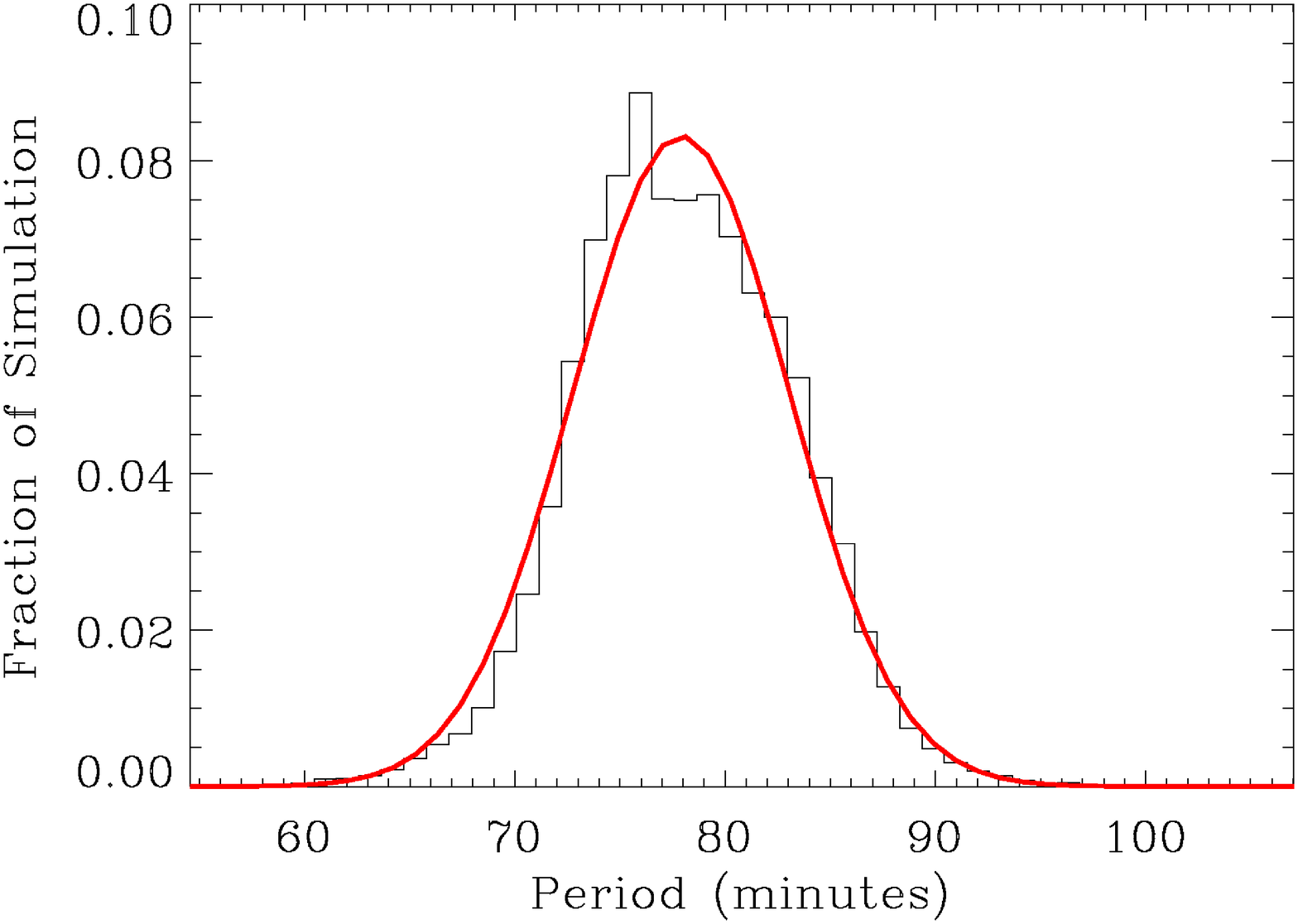}
	\end{subfigure}
	\quad
	\begin{subfigure}[t]{0.4\textwidth}
		\centering
		\includegraphics[width=\linewidth]{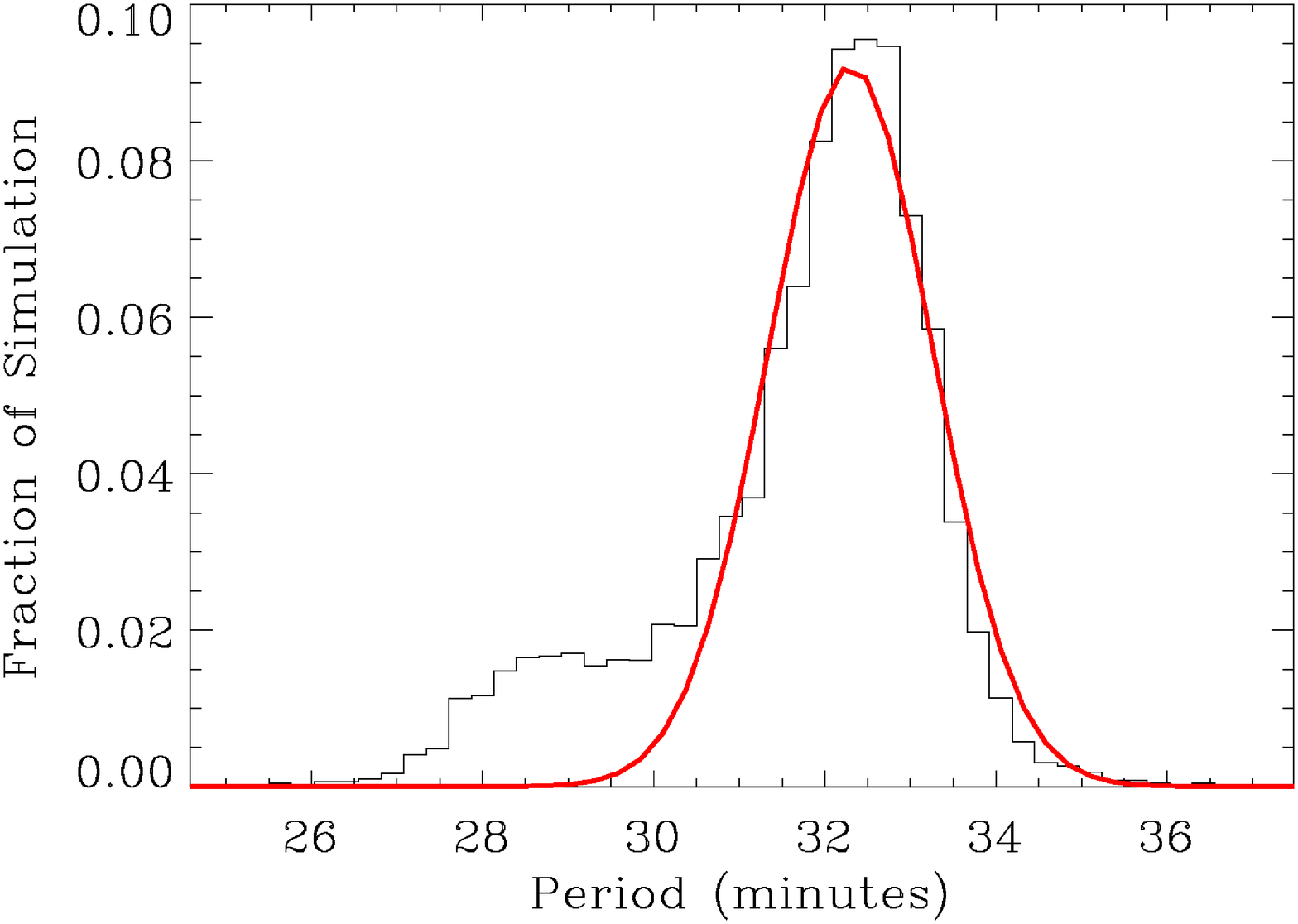}
	\end{subfigure}
	\caption{Histograms showing the results of Monte Carlo simulations for the main (left) and secondary (right) periods. The red overplotted curves show Gaussian fits, which have been used to estimate values of $78 \pm 12$\,min for the longer period, and $32 \pm 2$\,min for the shorter period.}
	\label{fig5}
\end{figure*}

Figure \ref{fig4} shows the result of fitting all parameters simultaneously and performing 10,000 Monte Carlo simulations, using a fitting function combining Equation 1 with the two periodicities, described by Equation 2. Histograms for the two fitted periods are shown in Figure \ref{fig5}. These have been fitted with Gaussians to give values of $78 \pm 12$ and $32 \pm 2$\,min, which are in good agreement with the values obtained using the wavelet and autocorrelation methods. The same Monte Carlo simulations were used to evaluate correlations between the different fitted parameters. No strong correlations were found between the periods themselves (8\% Pearson's correlation), or between the periods and their corresponding decay times (Pearson's correlation of 19\% for the longer periodicity and 29\% for the shorter periodicity), suggesting that they are independent. Although the longer periodicity was found to be slightly correlated with the decay time of the flare itself ($t_0$ in Equation 1), with a Pearson's correlation of 50\%, over 99\% of the fitted values were within the range of $84^{+25}_{-19}$\,min indicated by the wavelet analysis, suggesting that the data de-trending has not significantly influenced the results.

The significance of finding a multi-period QPP is that it is a strong indication that MHD oscillations are the cause of the QPPs in this flare. Multiple periods are difficult to explain with the load/unload mechanisms whereas harmonics are a common feature of resonators, and different types of wave have different characteristic periods and damping times. There is, however, a possibility that the QPP signal detected in the flare is non-linear, and such a signal could readily be produced by a self-oscillation \citep{2012ApJ...761..134N}. A non-linear signal, for example a sawtooth wave, can be constructed by the superposition of different sine/cosine waves (a Fourier series expansion), hence its Fourier/wavelet spectrum would have multiple peaks. In this case the phase difference between the sinusoidal components at a particular point in time would be $2\pi$ less than the phase difference at a point one cycle of the fundamental harmonic ahead of the first point. On the other hand, if the signals belong to different MHD modes or their spatial harmonics, they may have phases disconnected from one another. To check whether the detected periodicities are time harmonics of the same non-linear signal, the phases of oscillation of the two periodicities were found by fitting Equation 2, with a period equal to 78\,min, to the de-trended light curve (shown in Figure \ref{fig3}a), and fitting the shorter periodicity to the de-trended light curve with the fitted longer periodicity subtracted (shown in Figure \ref{fig3}c). The phases were found to be $3.8 \pm 0.1$ radians for the longer periodicity, and $0.6 \pm 0.2$ radians for the shorter periodicity, giving a phase difference of $3.2 \pm 0.2$ radians at the time $t=0$. After one cycle of the longer periodicity the phase difference is $5.7 \pm 0.2$ radians, and after two cycles it is $14.6 \pm 0.2$ radians. Since these phase differences do not differ by a factor of $2\pi$, this suggests that the signal is linear and that the shorter periodicity is a spatial harmonic of the longer periodicity, or the result of a different MHD mode. Also, the decay time of the shorter periodicity is $77 \pm 29$\,min, compared to $80 \pm 12$\,min for the longer periodicity. If this shorter periodicity were a higher harmonic of a non-linear signal, it would be expected to decay faster than the fundamental harmonic. Indeed it is possible that the shorter periodicity decays faster than the main oscillation due to the associated uncertainties, but it is more likely the case that they are similar in duration.

Considering the ratio of the periods; for a uniform medium the fundamental and second harmonics might be expected to be a factor of two different, whereas here the periods have a ratio of $2.4 \pm 0.4$. While some MHD modes are dispersionless or weakly dispersive (e.g. torsional and longitudinal), others are highly dispersive (e.g. kink and sausage) and so the ratio of the periods of their spatial harmonics can be non-integer \citep{2009A&A...493..259I}. Also stratification of the plasma density due to gravity, along with the geometry of coronal loops, means that the plasma density and magnetic field strength vary along the loop, and hence it is most likely that the wave frequency does not scale linearly with the wavelength for spatial harmonics. Therefore this ratio is consistent with the findings in the previous paragraph, and the presence of spatial harmonics. While we cannot use this information to conclusively identify the mechanism behind the QPPs, sausage modes can be excluded, as their characteristic periods are much shorter than those detected here \citep{2012ApJ...761..134N}.

\section{SUMMARY}

QPPs in the light curve of a flare on KIC9655129 were found, which, when analysed with the wavelet and autocorrelation techniques, showed evidence of the coexistence of two significant periods of oscillation, namely $78 \pm 12$\,min and $32 \pm 2$\,min. These are consistent with the presence of two spatial harmonics due to the dispersive nature of guided magnetohydrodynamic waves. Further evidence suggesting that these periodicities are not components of a non-linear signal was found by comparing the phase differences and decay times. While it is possible that these oscillations could be instrumental or astrophysical artifacts, there is no evidence of periodicities less than several hours in the rest of the data. Multiple periods are much more likely to be associated with MHD wave mechanisms of QPPs, rather than load/unload mechanisms, so this is one step towards understanding the physics at play, and further suggests that the underlying physics in solar and stellar flares could be similar. It is also possible that one periodicity is due to a load/unload mechanism, and the other due to an MHD oscillation.

\acknowledgements

CEP \& VMN: This work was supported by the European Research Council under the \textit{SeismoSun} Research Project No. 321141. VMN acknowledges support from the STFC consolidated grant ST/L000733/1. A-MB thanks the Institute of Advanced Study, University of Warwick for their support. We would like to thank the \emph{Kepler} team for providing the data used in this paper, obtained from the Mikulski Archive for Space Telescopes (MAST). Funding for the \emph{Kepler} mission is provided by the NASA Science Mission directorate. Wavelet software was provided by C. Torrence and G. Compo and is available at \\ \url{http://paos.colorado.edu/research/wavelets/}.

\end{document}